\documentclass[11pt]{article}
\pdfoutput=1
\usepackage{fullpage}
\usepackage{xcolor}
\usepackage{cite}
\usepackage{graphicx}
\usepackage{tensor}
\usepackage{amsmath}
\usepackage{amssymb}
\usepackage{subcaption}
\usepackage[utf8x]{inputenc}
\usepackage[hidelinks]{hyperref}
\title{}
\date{}
\renewcommand{\vec}[1]{\mbox{\boldmath$ #1 $}}

\def\beq{\begin{equation}}
\def\eeq{\end{equation}}
\begin{document}
\bibliographystyle{utphys}

% Commands
\newcommand{\be}{\begin{equation}}
\newcommand{\ee}{\end{equation}}
\newcommand\n[1]{\textcolor{red}{(#1)}} %in-text notes
\newcommand{\diff}{\mathop{}\!\mathrm{d}}
\newcommand{\lb}{\left}
\newcommand{\rb}{\right}
\newcommand{\f}{\frac}
\newcommand{\pd}{\partial}
\newcommand{\tr}{\text{tr}}
\newcommand{\fdiff}{\mathcal{D}}
\newcommand{\im}{\text{im}}
\let\caron\v
\renewcommand{\v}{\mathbf}
\newcommand{\T}{\tensor}
\newcommand{\R}{\mathbb{R}}
\newcommand{\C}{\mathbb{C}}
\newcommand{\Z}{\mathbb{Z}}
\newcommand{\msbar}{\ensuremath{\overline{\text{MS}}}}
\newcommand{\DIS}{\ensuremath{\text{DIS}}}
\newcommand{\abar}{\ensuremath{\bar{\alpha}_S}}
\newcommand{\bb}{\ensuremath{\bar{\beta}_0}}
\newcommand{\rc}{\ensuremath{r_{\text{cut}}}}
\newcommand{\Nd}{\ensuremath{N_{\text{d.o.f.}}}}
\newcommand{\red}[1]{{\color{red} #1}}

\titlepage
\begin{flushright}
UUITP-02/24\\
\end{flushright}

\vspace*{0.5cm}

\begin{center}
{\bf \Large What can abelian gauge theories teach us about\\ kinematic algebras?}

\vspace*{1cm} 
\textsc{Kymani Armstrong-Williams$^a$\footnote{k.t.k.armstrong-williams@qmul.ac.uk}, Silvia Nagy$^b$\footnote{silvia.nagy@durham.ac.uk},
Chris D. White$^a$\footnote{christopher.white@qmul.ac.uk}, Sam Wikeley$^c$\footnote{sam.wikeley@physics.uu.se}} \\

\vspace*{0.5cm} $^a$ Centre for Theoretical Physics, Department of
Physics and Astronomy, \\
Queen Mary University of London, 327 Mile End
Road, London E1 4NS, UK\\

\vspace*{0.5cm} $^b$ Department of Mathematical Sciences, Durham University, Durham, DH1 3LE, UK\\

\vspace*{0.5cm} $^c$ Department of Physics and Astronomy, Uppsala University, Box 516, 75120 Uppsala, Sweden\\

\end{center}

\vspace*{0.5cm}

\begin{abstract}
  The phenomenon of BCJ duality implies that gauge theories possess an
  abstract kinematic algebra, mirroring the non-abelian Lie algebra
  underlying the colour information. Although the nature of the
  kinematic algebra is known in certain cases, a full understanding is
  missing for arbitrary non-abelian gauge theories, such that one
  typically works outwards from well-known examples. In this paper, we
  pursue an orthogonal approach, and argue that simpler abelian gauge
  theories can be used as a testing ground for clarifying our
  understanding of kinematic algebras. We first describe how classes
  of abelian gauge fields are associated with well-defined subgroups of the diffeomorphism algebra. By considering certain special subgroups, we show that one may construct
  interacting theories, whose kinematic algebras are inherited from
  those already appearing in a related abelian theory. Known
  properties of (anti-)self-dual Yang-Mills theory arise in this way,
  but so do new generalisations, including self-dual electromagnetism
  coupled to scalar matter. Furthermore, a recently obtained
  non-abelian generalisation of the Navier-Stokes equation fits into
  a similar scheme, as does Chern-Simons theory. Our results provide useful input to further
  conceptual studies of kinematic algebras.
\end{abstract}

% \vspace*{0.5cm}
\newpage

\tableofcontents

\section{Introduction}
\label{sec:intro}

(Non-)abelian gauge theories continue to be intensely studied, due to
their role in describing our universe at its most fundamental level,
as accessible in current experiments. In 2008, a remarkable new
structure was discovered in the scattering amplitudes of non-abelian
theories, which has become known as {\it BCJ
  duality}\cite{Bern:2008qj}. In simple terms, it states that certain
kinematic parts of scattering amplitudes (i.e. which depend on
momentum and / or polarisation information) can be made to obey
similar identities to those satisfied by the colour charge
information. Given that the latter constraints arise from the Lie
algebra underlying the gauge symmetry of the theory, it appears
to be the case that gauge theories have some sort of {\it kinematic
  algebra}, which had previously remained hidden. Quite what this new
structure is trying to tell us -- and how far-reaching its ultimate
scope and consequences are -- remain mysterious, not least due to the
fact that we are ignorant of what the kinematic algebra actually
is, for most non-abelian gauge theories of physical interest. It does
not help that the BCJ duality property only usually shows up
iteratively order-by-order in perturbation theory, so that it is not
understood in a fully non-perturbative way. Nevertheless, one very
striking consequence of BCJ duality is already known: provided gauge
theory amplitudes are written so that the kinematic algebra is
manifest ({\it BCJ-dual form}), one may simply replace colour
information by supplementary kinematic factors, as well as coupling
constants, to obtain amplitudes in gravity theories. This is known as
the {\it double copy}~\cite{Bern:2010ue,Bern:2010yg}, and is motivated
by earlier work in string theory~\cite{Kawai:1985xq}. In the field
theory context, the double copy has been extended to classical
solutions~\cite{Monteiro:2014cda,Luna:2015paa,Luna:2016due,Luna:2016hge,Luna:2017dtq,Bahjat-Abbas:2017htu,Luna:2018dpt,Goldberger:2016iau,Goldberger:2017frp,Goldberger:2017ogt},
leading to practical applications such as new techniques for
gravitational wave physics (see
e.g. refs.~\cite{Bern:2022wqg,Adamo:2022dcm,Kosower:2022yvp,Buonanno:2022pgc}
for recent reviews). A parallel research frontier examines conceptual
questions raised by BCJ duality and the double copy, both of
which promise tantalising hints that our traditional ways of thinking
about field theories might have obscured a deep underlying
structure. We should thus leave no stone unturned in performing case
studies relating to this structure, including looking at aspects of
BCJ duality that may have been previously overlooked.

The first concrete example of a kinematic algebra being found for a
(partial) theory was the case of self-dual Yang-Mills theory, in the
by now well-known analysis of ref.~\cite{Monteiro:2011pc}. This theory
corresponds to keeping only one of the (circular) polarisation states
of the gluon, and a convenient language for this theory exists in a
particular choice of gauge (the lightcone gauge), such that the
Lagrangian for the theory is manifestly cubic~\cite{Parkes:1992rz}. As
the authors of ref.~\cite{Monteiro:2011pc} showed, the cubic vertex
can be seen to contain two sets of structure constants, one of which
corresponds to the known non-abelian gauge group. The other structure
constants correspond to an area-preserving diffeomorphism algebra,
which is infinite-dimensional, and the fact that the structure
constants appear alongside each other in the single interaction term
for the theory means that any perturbative solutions will necessarily
obey BCJ duality. Furthermore, a known equation for self-dual gravity
(the {\it Plebanski equation}~\cite{Plebanski:1975wn}) follows
straightforwardly upon replacing the colour structure constants with a
second set of kinematic ones. Deformations of both self-dual gauge /
gravity theories, giving rise to generalisations of these structure
constants, have been presented in
refs.~\cite{BjerrumBohr:2012mg,Chacon:2020fmr,Lipstein:2023pih}. 

Another case in which the kinematic algebra is known is that of
Chern-Simons theory in three spacetime dimensions, as examined in
ref.~\cite{Ben-Shahar:2021zww}. The authors worked in Lorenz gauge,
and showed that the cubic interaction vertex of the theory can be
associated with the structure constants of a volume-preserving
diffeomorphism algebra. Interestingly, this conclusion extended beyond
the physical field itself. Working in Lorenz gauge necessitates the
introduction of Faddeev-Popov ghost fields, whose role is to subtract
the unphysical degree of freedom carried by the off-shell gauge
field. The ghost fields can be combined with the gauge field to make a
``superfield'' living in superspace, where the latter possesses
anti-commuting coordinates in addition to the spacetime
ones~\cite{Axelrod:1991vq,Axelrod:1993wr}. Using this formalism,
ref.~\cite{Ben-Shahar:2021zww} showed that the full superfield 3-vertex
gives rise to an extended volume-preserving diffeomorphism algebra,
where the ``volume'' is separately preserved in the subspaces of
(anti-)commuting coordinates. Both self-dual Yang-Mills and Chern-Simons theory (in Lorenz gauge) were argued to be special cases of a more general theory dubbed {\it semi-abelian Yang-Mills theory} in ref.~\cite{Edison:2023ulf}. We will revisit this theory in what follows.

The idea that off-shell degrees of freedom need to be explicitly
included in kinematic algebras has been taken further in
e.g. refs.~\cite{Reiterer:2019dys,Borsten:2020zgj,Borsten:2021hua,Borsten:2021gyl,Borsten:2022ouu,Borsten:2022vtg,Borsten:2023ned,Diaz-Jaramillo:2021wtl,Bonezzi:2022yuh,Bonezzi:2022bse,Bonezzi:2023ciu}, 
which also consider the idea that the latter may not be conventional
Lie algebras. A Lie algebra is characterised by a vector space $V$ and
a {\it Lie bracket}: $V\times V\rightarrow V$, that takes a pair of
elements of $V$, and associates this with a third element. This
bracket satisfies the well-known Jacobi identity, and the above
references argue that this structure is insufficient to describe
arbitrary field theories. Rather, these are expected to be built upon
so-called $L_\infty$ or {\it strong homotopy} algebras. These can be
viewed as generalisations of Lie algebras, where the Jacobi identity
is satisfied only up to terms involving higher-order brackets. Each type of bracket obeys
further identities that are satisfied only up to terms involving yet
higher-order brackets, resulting in an intricate structure of
constraints. A particular homotopy algebra known as a
$\text{BV}_\infty^\Box$ algebra has been shown to be relevant for full
Yang-Mills theory~\cite{Reiterer:2019dys}, itself a generalisation of
the well-known Batalin-Vilkovisky formalism for gauge
theory~\cite{Batalin:1981jr,Batalin:1983ggl}. Reference~\cite{Bonezzi:2022bse}
showed how the kinematic algebra for Chern-Simons theory, previously
identified in ref.~\cite{Ben-Shahar:2021zww}, could be cast into this
framework. More recently, ref.~\cite{Bonezzi:2023pox} reexamined the
case of self-dual Yang-Mills theory using the $\text{BV}_\infty^\Box$
ideas, in particular addressing the question of whether the
straightforward Lie algebra of area-preserving diffeomorphisms found
for lightcone gauge in ref.~\cite{Monteiro:2011pc} extends to more
general gauges. The
conclusion was that, in general, it is not expected that the
$\text{BV}_\infty^\Box$ algebra reduces to a Lie algebra, even in the
self-dual sector. Further work on kinematic algebras in a variety of
contexts can be found in
refs.~\cite{Mizera:2019blq,Fu:2016plh,Chen:2019ywi,Chen:2022nei,Brandhuber:2021bsf,Brandhuber:2022enp,Mafra:2014oia,Mafra:2015vca}.

Despite (or perhaps because of!) the above progress, a number of open
questions remain: how do we find the kinematic algebras of particular
theories, either as reductions of $L_\infty$ algebras or otherwise?
Are kinematic algebras gauge-dependent in general? If so, is there an
optimal choice of gauge, such that the kinematic algebra is somehow
minimal? Is it always possible to reduce it to a Lie algebra? In this
paper, we will explore some of these questions in the context of
simple abelian gauge theories, and our motivations are as follows. For
starters, it is often claimed -- erroneously -- that there is no
manifestation of BCJ duality for linear (or linearised) gauge
theories. That this is not in fact true rests on the fact that one may
indeed associate a ``kinematic algebra'' with linear gauge fields,
given that they are Lie-algebra valued in two different Lie
algebras. The first of these is the usual gauge algebra of the theory,
and the second is the algebra of diffeomorphisms generated by the
(vector) gauge field. For self-dual linearised solutions that can be double-copied to make gravity solutions, one
must replace the colour generators with a second set of diffeomorphism
generators, which indeed amounts to a colour-kinematics duality, as
argued in ref.~\cite{Armstrong-Williams:2022apo}. In this paper, we
wish to expand upon and clarify this point, by defining more precisely
the above ideas, which were only briefly introduced in
ref.~\cite{Armstrong-Williams:2022apo}. In particular, we will see
that certain gauge choices and / or solution types in abelian gauge
theory pick out well-defined subgroups of the full diffeomorphism
group, such that known cases of kinematic algebras correspond to some
of these subgroups.

Admittedly, the diffeomorphism algebras that arise at linear level are
not what people usually mean when they talk about kinematic algebras,
which are instead associated with interactions between
fields. However, we can then use abelian gauge theories to clarify
aspects of more general kinematic algebras. Given that any interacting
theory (including a non-abelian gauge theory) must have a
non-interacting linearisation, we can ask which of our ``special''
subgroups of diffeomorphisms can be preserved by the inclusion of
interactions. We will see that a particularly interesting case is when
the gauge field generates so-called {\it symplectomorphisms} or, in
other words, when the field itself is {\it Hamiltonian}. We will
review the definition of these concepts below, but the presence of a
Hamiltonian vector field allows us to define a kinematic Poisson
bracket, which can in turn be used to construct interacting theories
that contain a non-trivial kinematic algebra. We will show that the
known kinematic algebra of (anti-)self dual Yang-Mills theory in
lightcone gauge arises as a special case of this construction, but
that there are also various generalisations of this story. As a novel
byproduct, we will also see that the self-dual sector of QED coupled
to scalar matter has a straightforward kinematic algebra, in terms of a similar Poisson bracket. There are also cases in which the Lie bracket of diffeomorphisms itself arises as part of an interaction term, and we will show that both Chern-Simons theory~\cite{Ben-Shahar:2021zww} and a recent non-abelian generalisation of the Navier-Stokes equation~\cite{Cheung:2020djz} in three spacetime dimensions arise in this way. In all cases, there is a geometric understanding of the kinematic
algebra, in that it corresponds to the diffeomorphisms generated by
the gauge field.  We hope that our results are useful for
further studies of such algebras, including guiding searches for
higher geometric structures upon which they act.

The structure of our paper is as follows. In
section~\ref{sec:abelian}, we review and expand the ideas of
ref.~\cite{Armstrong-Williams:2022apo}, showing how sectors of
linearised gauge theories can be classified according to their
diffeomorphism algebras. In section~\ref{sec:interactions}, we
describe how the kinematic algebras of interacting theories can be
built upon the subgroups of diffeomorphisms one encounters at
linearised level, giving a number of examples, some previously
unknown. We discuss our results and conclude in
section~\ref{sec:discuss}.

\section{Diffeomorphisms and linearised gauge theories}
\label{sec:abelian}

\subsection{Linearised self-dual fields}
\label{sec:self-dual}

\begin{figure}
  \begin{center}
    \scalebox{0.6}{\includegraphics{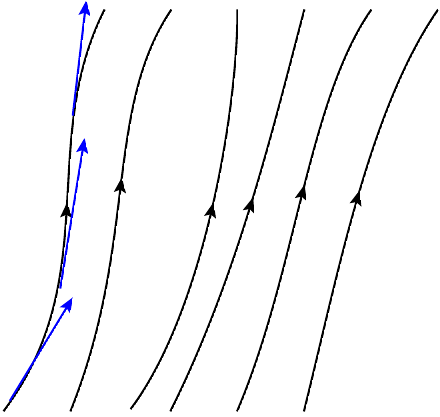}}
    \caption{A vector field $V^\mu$ (shown in blue) generates integral
      curves, such that $V^\mu(x)$ is tangent to the integral curve
      passing through $x^\mu$. }
    \label{fig:curves}
  \end{center}
\end{figure}

Reference~\cite{Armstrong-Williams:2022apo} examined solutions of
linearised non-abelian gauge theories, whose field equations constrain
a field
\begin{equation}
  {\bf A}_\mu=A_\mu^a {\bf T}^a.
  \label{Amudef}
\end{equation}
Here Greek and Latin letters specify spacetime and adjoint (colour)
indices respectively, and ${\bf T}^a$ is a generator of the gauge
group. The set of all generators satisfies the Lie algebra
\begin{equation}
  [{\bf T}^a,{\bf T}^b]=if^{abc}{\bf T}^c,
  \label{Lie}
\end{equation}
with structure constants $f^{abc}$, such that the gauge field itself
takes values in the Lie algebra. As pointed out in
ref.~\cite{Armstrong-Williams:2022apo} (see also
ref.~\cite{Fu:2016plh}), the field ${\bf A}_\mu$ is in fact valued in
a second Lie algebra. To see this, we may recall that a given vector
field $V^\mu$ on a manifold generates infinitesimal
diffeomorphisms
\begin{equation}
  V^\mu(x)\partial_\mu,
  \label{diffeo}
\end{equation}
which can be visualised geometrically as follows. First, one may
construct the {\it integral curves} (fieldlines) of the vector field
$V^\mu$. These are an infinite set of non-intersecting curves, such
that $V^\mu(x)$ is tangent to the integral curve passing through
$x^\mu$. An example of these integral curves is shown in
figure~\ref{fig:curves}, and the action of eq.~(\ref{diffeo}) is to
effect an infinitesimal translation along each curve
simultaneously. The set of all vector fields on a manifold then
consists of the set of all such diffeomorphisms, and they form an
algebra under the {\it Lie bracket}
\begin{equation}
  [V^{(1)\mu}\partial_\mu,V^{(2)\nu}\partial_\nu]=
  V^{(3)\mu}\partial_\mu.
  \label{Lie2}
\end{equation}
As in eq.~(\ref{Lie}), one derives the right-hand side by forming the
commutator of the two transformations on the left-hand side, and the
algebra is closed given that the bracket of two vector fields is
itself a vector field. The components of the latter turn out to be
given by
\begin{equation}
  V^{(3)\mu}=V^{(1)\nu}\partial_\nu V^{(2)\mu}
  -V^{(2)\nu}\partial_\nu V^{(1)\mu}.
  \label{V3def}
\end{equation}
The Lie bracket has a geometric interpretation as the {\it Lie
  derivative} of the vector field $V^{(1)\mu}$ along the vector field
$V^{(2)\mu}$, and one may also interpret eq.~(\ref{V3def}) as representing
the failure of a loop made of infinitesimal diffeomorphisms along two
different vector fields to close in general. 

Now let us consider an abelian-like solution of a non-abelian gauge
theory, for which one may make the ansatz
\begin{equation}
  A_\mu^a=c^a A_\mu,
  \label{Amuabel}
\end{equation}
for constant colour vector $c^a$, as has been done in the context of
the double copy in e.g. ref.~\cite{Monteiro:2014cda}. One may then
consider $A_\mu$ to be a solution of an abelian gauge theory, and it
will generate diffeomorphisms as described above. There are several
known cases in which abelian-like gauge fields can be double-copied to
make gravity solutions. An example relevant for the present paper is
the case of self-dual linearised fields in lightcone gauge, examined in
ref.~\cite{Armstrong-Williams:2022apo}, and which are such that the
gauge field can be written as
\begin{equation}
  A_\mu=\hat{k}_\mu\phi,
  \label{Amukhat}
\end{equation}
where $\phi(x)$ is a scalar field, and $\hat{k}$ a differential
operator satisfying\footnote{We use $A\cdot B$ to denote $A^\rho B_\rho$, where $A$ and $B$ could be either fields or operators.}
\begin{equation}
  \hat{k}^2=0,\quad \partial\cdot\hat{k}=0.
  \label{khat}
\end{equation}
The general form of such an operator was found in
ref.~\cite{Armstrong-Williams:2022apo} to be expressible (in Euclidean
spacetime signature) as\footnote{We do not raise or lower indices in
eq.~(\ref{khatdef}) and subsequent equations, as a reminder that we
are in Euclidean signature.}
\begin{equation}
  \hat{k}_\mu =B_i\bar{\eta}_{\mu\nu}^i\partial_\nu,
  \label{khatdef}
\end{equation}
where
\begin{equation}
  \bar{\eta}_{\mu\nu}^1=\left(
  \begin{array}{cccc}
    0 & 0 & 0 & -1\\
    0 & 0 & 1 & 0\\
    0 & -1 & 0 & 0\\
    1 & 0 & 0 & 0\\
  \end{array}
  \right),\quad
  \bar{\eta}_{\mu\nu}^2=\left(
  \begin{array}{cccc}
    0 & 0 & -1 & 0\\
    0 & 0 & 0 & -1\\
    1 & 0 & 0 & 0\\
    0 & 1 & 0 & 0\\
  \end{array}
  \right),\quad
  \bar{\eta}_{\mu\nu}^3=\left(
  \begin{array}{cccc}
    0 & 1 & 0 & 0\\
    -1 & 0 & 0 & 0\\
    0 & 0 & 0 & -1\\
    0 & 0 & 1 & 0\\
  \end{array}
  \right)
  \label{tHooft1}
\end{equation}
are so-called {\it 't Hooft symbols}, which arise in the study of
non-abelian (self-dual) instantons, and $B_i$ a constant vector such that $\vec{B}^2=0$. Then, the field
\begin{equation}
  h_{\mu\nu}=\hat{k}_\mu\hat{k}_\nu \phi
  \label{hmunuSD}
\end{equation}
is a self-dual gravity solution. A canonical example of this
construction is the Eguchi-Hanson gravitational instanton, first
considered from a double-copy point of view in
ref.~\cite{Berman:2018hwd} (see also ref.~\cite{Luna:2018dpt}). We
usually think of the double copy as replacing a colour Lie algebra
with a second copy of the kinematic algebra. Thus,
ref.~\cite{Armstrong-Williams:2022apo} suggested identifying the Lie
algebra of diffeomorphisms with the ``kinematic algebra'' of an
abelian gauge field. It should be stressed again that this is not
usually what we mean when we talk about kinematic algebras, which are
instead associated with interaction terms in a theory. Hence, we shall
continue to refer to the Lie algebra generated by abelian solutions as
a diffeomorphism algebra (which of course it is) in what
follows. Nevertheless, the particular diffeomorphisms generated by the
self-dual abelian solutions of eq.~(\ref{Amukhat}) have a particularly
elegant geometric interpretation. One may express the gauge field of
eqs.~(\ref{Amukhat}, \ref{khatdef}) as 
\begin{equation}
  A_{\mu}\partial_{\mu}=(\hat{k}_\mu\phi)\partial_\mu=\left(b^{(1)}_{[\mu}b^{(2)}_{\nu]}\partial_\nu
  \phi\right)\partial_\mu,
  \label{Bidecomp}
\end{equation}
with\footnote{We have assumed $B_1\neq 0$ in eq.~(\ref{bmu12}), but
similar solutions can be derived for $B_1=0$.}
\begin{equation}
  b_\mu^{(1)}=(B_1,B_2,B_3,0),\quad b_\mu^{(2)}=\left(0,\frac{B_3}{B_1},
  -\frac{B_2}{B_1},-1\right).
  \label{bmu12}
\end{equation}
It then follows that the diffeomorphisms generated by
eq.~(\ref{Bidecomp}) take place in the family of null planes whose
tangent bivectors are constructed from $b^{(1)}_\mu$ and $b^{(2)}_\mu$
(but which may have some displacement from the origin). Here the null property arises from the fact that
\begin{equation}
    b^{(i)}\cdot b^{(j)} = 0, \quad \forall i \in \{1,2\}.
\end{equation}
Given also the
condition $\partial\cdot \hat{k}=0$, these diffeomorphisms will be
area-preserving, such that the diffeomorphism algebra of self-dual
abelian solutions is that of area-preserving diffeomorphisms.

As mentioned in the introduction, an area-preserving diffeomorphism
algebra also arises in self-dual Yang-Mills theory in lightcone gauge,
even when nonlinear interactions are included. Indeed, it is the {\it
  same} area-preserving diffeomorphism algebra as has been found in
the case of abelian fields discussed above. One clue as to why this
happens can be found in ref.~\cite{Monteiro:2014cda}, which showed
that substituting the fully non-abelian ansatz
\begin{equation}
  A_\mu^a=\hat{k}_\mu \Phi^a
  \label{Amuakhat}
\end{equation}
into the Yang-Mills equations, where $\hat{k}_\mu$ satisfies the
conditions of eq.~(\ref{khat}), leads to the known equation of
self-dual Yang-Mills theory in lightcone gauge, whose underlying
area-preserving diffeomorphism algebra was discovered in
ref.~\cite{Monteiro:2011pc}. This suggests that the kinematic algebra
of self-dual Yang-Mills theory -- a bona fide kinematic algebra of an
interacting theory -- is somehow related to the self-dual
diffeomorphisms found in the abelian theory. In what follows, we will
fully explain this connection, for which we first need to discuss
abelian diffeomorphisms in more detail.

\subsection{The space of abelian diffeomorphisms}
\label{sec:abeliandiff}

\begin{figure}
  \begin{center}
    \scalebox{1.0}{\includegraphics{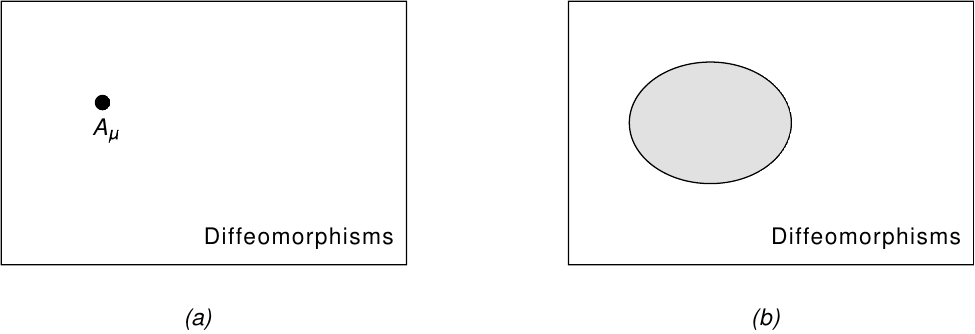}}
    \caption{(a) A given gauge field $A_\mu$ constitutes a point in
      the space of all possible diffeomorphisms in spacetime; (b)
      closed subgroups of the general diffeomorphism algebra form
      well-defined blobs in the space of all possible diffeomorphisms,
      which can in turn be related to a special class of abelian gauge
      fields. }
    \label{fig:diffplot}
  \end{center}
\end{figure}

A given abelian gauge field $A_\mu$ generates diffeomorphisms along a
particular set of integral curves. We can represent this pictorially
as in figure~\ref{fig:diffplot}(a), which shows the set of all possible
diffeomorphisms. Our given gauge field is then a point in this
diagram. Note that there is no non-abelian algebra associated with
this point: by eqs.~(\ref{Lie2}, \ref{V3def}), the diffeomorphisms
associated with $A_\mu$ are mutually commuting. This makes sense from
figure~\ref{fig:curves}, given that performing one simultaneous
translation along all possible integral curves, followed by another,
is clearly insensitive to the order in which these translations are
carried out.

Non-trivial diffeomorphism algebras will consist of subgroups of the
full diffeomorphism algebra, which can be represented pictorially as
in figure~\ref{fig:diffplot}(b). From figure~\ref{fig:diffplot}, each
point inside such a closed subgroup will correspond to an abelian
gauge field $A_\mu$, and we can therefore ask the following question:
given a particular subgroup of diffeomorphisms, can we identify a
particular class of abelian gauge fields that this corresponds to?
There are in fact two particular subgroups of the (infinitely
dimensional) diffeomorphism group, whose physical interpretation is
easy to appreciate.

\subsubsection{Volume-preserving diffeomorphisms}
\label{sec:volume}

It is well-known that volume-preserving diffeomorphisms form a closed
subgroup. Furthermore, a simple criterion for a diffeomorphism to be
volume preserving is that the (multidimensional) divergence of the
vector field vanishes:
\begin{equation}
  \partial\cdot A=0.
  \label{Lorenz}
\end{equation}
For an abelian gauge field, this is the Lorenz gauge condition. Thus,
abelian gauge fields in the Lorenz gauge generate a volume-preserving
diffeomorphism algebra. Interestingly, ref.~\cite{Ben-Shahar:2021zww}
studied Chern-Simons theory in Lorenz gauge, finding that the
kinematic algebra of the theory is indeed that of volume-preserving
diffeomorphisms. As in the case of self-dual Yang-Mills theory, the
kinematic algebra of Chern-Simons is a statement about the interaction
vertex. However, the fact that it parallels the diffeomorphism algebra
that already exists at linear level is reminiscent of how the
kinematic algebra of self-dual Yang-Mills (area-preserving
diffeomorphisms) is closely related to the diffeomorphism algebra of
self-dual abelian solutions, in lightcone gauge.

Subgroups of the full spacetime volume-preserving diffeomorphism group
also exist. In particular, one may consider taking a fixed hypervolume
of lower dimension than the total spacetime dimension. Then,
diffeomorphisms that preserve this lower-dimensional volume form a group by themselves. An example of this is the group of area-preserving
diffeomorphisms seen in self-dual Yang-Mills theory.

\subsubsection{Symplectomorphisms}
\label{sec:symplect}

A {\it symplectic manifold} is a manifold that is endowed with a
particular two-form called the {\it symplectic form}, and whose
existence allows us to define additional structures. A familiar
example of a symplectic manifold in classical point particle mechanics
is that of the phase space of a system, consisting of a set of
generalised coordinates $\{q_i\}$, and momenta $\{p_i\}$. For $N$
position degrees of freedom, we may then write the total set of $2N$
phase space coordinates as $\{\xi_i\}\equiv\{q_i,p_i\}$, and the
symplectic form is given by 
\begin{equation}
  \omega=\sum_i dq_i\wedge dp_i=\Omega_{ij}d\xi_i\wedge d\xi_j,
  \label{omegadef}
\end{equation}
where $\Omega_{ij}$ consists of the $2N\times 2N$ matrix
\begin{equation}
  \Omega_{ij}=\left(\begin{array}{cc}{\bf 0} & {\bf I}\\
    -{\bf I} & {\bf 0}\end{array}\right),
  \label{Omegaij}
\end{equation}
and we have used the shorthand notation ${\bf 0}$ and ${\bf I}$ for an
$N\times N$ zero or identity matrix respectively. Vector fields can be
defined on phase space, whose integral curves associate particular
values of the positions $\{q_i\}$ and momenta $\{p_i\}$ at any given
value of some parameter. If we interpret this parameter as the time,
then these integral curves represent possible histories of a classical
system which, for a second-order equation of motion, can indeed be
entirely specified by describing how the positions and momenta evolve
as time progresses.

A {\it Hamiltonian vector field} is a vector field that preserves the
symplectic form. Such fields can be written in general as
\begin{equation}
  V_i=\Omega_{ij}\partial_j H(\{\xi_i\}),
  \label{Videf}
\end{equation}
where $\partial_j$ represents the partial derivative in the full set
of generalised coordinates and momenta $\{\xi_j\}$. The function $H$
is called the {\it Hamiltonian}, and governs the time evolution of the
system, in that it controls the diffeomorphisms along the vector field
$V_i$, which we have already stated represents all possible
histories. More precisely, the fact that $V_i$ is tangent to an
integral curve parametrised by time $t$ implies
\begin{equation}
  V_i=\frac{d\xi_i}{dt},
  \label{Vidt}
\end{equation}
such that the diffeomorphisms generated by $V_i$ are of the form
\begin{equation}
  V_i\partial_i=\frac{d\xi_i}{dt}\partial_i\equiv \frac{d}{dt}.
  \label{Vidt2}
\end{equation}
This shows that $V_i$ generates time translations along each integral
curve as required, and eq.~(\ref{Videf}) now yields
\begin{equation}
  \frac{d}{dt}=\Omega_{ij}(\partial_j H)\partial_i.
    \label{dtH}
\end{equation}
In writing the equations of motion of the system, we may introduce the
{\it Poisson bracket}, which is formally defined through the action of the symplectic form on two Hamiltonian vector fields. Denoting the latter by
\begin{equation}
X_f=\Omega_{ij}(\partial_j f)\partial_i,\quad
X_g=\Omega_{ij}(\partial_j g)\partial_i
    \label{XfXg}
\end{equation}
for two scalar functions $f$ and $g$, one then has
\begin{equation}
\{f,g\}=\omega(X_f,X_g)=\Omega_{ij}(\Omega_{ik}\partial_k f) 
(\Omega_{jl}\partial_l g).
    \label{Poissondef}
\end{equation}
The right-hand side simplifies upon noticing that the symplectic form coefficients of eq.~(\ref{Omegaij}) satisfy 
\begin{equation}
    \Omega_{ij}\Omega_{ik}=\delta_{jk},
    \label{Omegarel}
\end{equation}
such that eq.~(\ref{Poissondef}) becomes
\begin{equation}
  \{f,g\}\equiv\Omega_{ij}(\partial_i f)(\partial_j g)=
  \frac{\partial f}{\partial q_i}\frac{\partial g}{\partial p_i}
  -  \frac{\partial g}{\partial q_i}\frac{\partial f}{\partial p_i}.
  \label{Poisson}
\end{equation}
Equation~(\ref{dtH}) then implies
\begin{equation}
  \frac{d\xi_i}{dt}=\{\xi_i,H\},
  \label{Hamilton1}
\end{equation}
or
\begin{equation}
  \frac{dq_i}{dt}=\frac{\partial H}{\partial p_i},\quad
  \frac{dp_i}{dt}=-\frac{\partial H}{\partial q_i},
  \label{Hamilton2}
\end{equation}
which we recognise as {\it Hamilton's equations} of classical mechanics.

Although the Hamiltonian formalism provides arguably the most
frequently encountered application of symplectic manifolds -- at least
for the physicist -- the language of Hamiltonian vector fields and
Poisson brackets is used whenever a manifold is equipped with a
symplectic form. Returning to the case of abelian gauge fields in
Euclidean signature, we can define a symplectic form
\begin{equation}
  \omega=\Omega_{\mu\nu}dx_\mu\wedge dx_\nu.
  \label{symplectic2}
\end{equation}
By definition, a symplectic form must be non-degenerate, meaning that the components $\Omega_{\mu\nu}$ are those of a non-singular matrix.  
We may then consider the special class of Hamiltonian gauge fields,
which by analogy with eq.~(\ref{Videf}) are given by
\begin{equation}
  A_\mu=\Omega_{\mu\nu}\partial_\nu \phi,
  \label{AOmega}
\end{equation}
for some scalar field $\phi$. The definition of such fields is that they preserve the symplectic form, which amounts to the condition that the Lie derivative of $\omega$ along a Hamiltonian vector field $A$ is zero. In components this condition reads (see e.g. ref.~\cite{Nakahara:2003nw})
\begin{equation}
    {\cal L}_A\,\omega=(A_\rho\partial_\rho\Omega_{\mu\nu}
    +\Omega_{\mu\rho}\partial_\nu A_\rho
    +\Omega_{\rho\nu} \partial_\mu A_\rho)dx_\mu\wedge dx_\nu=0.
    \label{Lieomega1}
\end{equation}
Upon taking the coefficients $\{\Omega_{\mu\nu}\}$ to be constant and using eq.~(\ref{AOmega}), eq.~(\ref{Lieomega1}) implies
\begin{equation}
\left(\Omega_{\mu\rho}\Omega_{\rho\alpha}\partial_\alpha\partial_\nu\phi
+\Omega_{\rho\nu}\Omega_{\rho\alpha}\partial_\mu\partial_\alpha\phi
\right)
dx_\mu\wedge dx_\nu
=0.
\label{Lieomega2}
\end{equation}
By analogy with eq.~(\ref{Omegarel}), we may take  
\begin{equation}
\Omega_{\rho\mu}\Omega_{\rho\alpha}=\delta_{\mu\alpha},
\label{Omegacond1}
\end{equation}
which indeed satisfies eq.~(\ref{Lieomega2}). We may also introduce a Poisson bracket (c.f. eq.~(\ref{Poissondef})):
\begin{equation}
\{\phi_1,\phi_2\}=\Omega_{\mu\nu}(\Omega_{\mu\alpha}\partial_\alpha\phi_1)
(\Omega_{\nu\beta}\partial_\beta\phi_2)=
\Omega_{\mu\nu}(\partial_\mu\phi_1)(\partial_\nu\phi_2),
\label{PoissonA}
\end{equation}
where the second equality follows from eq.~(\ref{Omegacond1}). 

The diffeomorphism generated by a Hamiltonian vector field is called a
{\it symplectomorphism}, and to show that such transformations form a
well-defined subgroup of the full diffeomorphism algebra, we must
verify that the Lie bracket of two Hamiltonian gauge fields is itself
Hamiltonian. This is indeed true, such that we may write
\begin{equation}
  [A^{(1)}_\mu\partial_\mu,A^{(2)}_\nu\partial_\nu]
  =A^{(3)}_\mu\partial_\mu,
  \label{Asymp}
\end{equation}
where all $A^{(i)}_\mu$ are Hamiltonian:
\begin{equation}
  A_\mu^{(i)}=\Omega_{\mu\nu}\partial_\nu\phi_i.
  \label{AiOmega}
\end{equation}
A standard result of symplectic geometry then states that the
``Hamiltonian'' $\phi_3$ on the right-hand side of eq.~(\ref{Asymp})
is related to the Poisson bracket of two Hamiltonians on the left-hand
side:
\begin{equation}
  \phi_3=-\{\phi_1,\phi_2\},
\label{phi3def}
\end{equation}
and we will make use of this result later on. 

As for volume-preserving diffeomorphisms, one may consider closed subgroups of the symplectomorphism subgroup. Given that the space that symplectomorphisms act upon must be even-dimensional, there is only one additional possibility in four spacetime dimensions. That is, one may take a two-dimensional subspace of four-dimensional spacetime, and define a symplectic form in this space. Let us denote the relevant symplectic form coefficients by $\omega_{ij}$, where the indices $i,j\in\{1,2\}$ span the independent coordinates in this subspace. The symplectic form coefficients are antisymmetric, and we can fix the normalisation such that these are equal to the two-dimensional Levi-Civita tensor:
\begin{equation}
\omega_{ij}=\epsilon_{ij},
\label{omegaeps}
\end{equation}
which is the two-dimensional analogue of eq.~(\ref{Omegaij}). By analogy with eq.~(\ref{Poisson}), the Poisson bracket of two scalar fields will be given by 
\begin{equation}
    \{\phi_1,\phi_2\}=\omega_{ij}(\partial_i\phi_1)(\partial_j\phi_2).
    \label{Poisson2d}
\end{equation}
where the difference with respect to eq.~(\ref{PoissonA}) is that the indices run only over the coordinates of the two-dimensional subspace acted on by symplectomorphisms, rather than the full four-dimensional spacetime volume. In what follows, it will nevertheless be convenient to work with 4-dimensional covariant notation, in which a Hamiltonian vector field defined with respect to a two-dimensional symplectic form is written as in eq.~(\ref{AOmega}). In that case, the coefficients $\Omega_{\mu\nu}$ contain $\Omega_{ij}$ as a submatrix, and the former do not strictly constitute the coefficients of a symplectic form due to the fact that the matrix $\Omega_{\mu\nu}$ is singular. However, it is straightforward to verify that 
\begin{equation}
    \Omega_{\mu\nu}(\partial_\mu\phi_1)(\partial_\nu\phi_2)
    =\omega_{ij}(\partial_i\phi_1)(\partial_j\phi_2).
    \label{Poissonrel}
\end{equation}
That is, one may continue to use the final expression in eq.~(\ref{PoissonA}) for the Poisson bracket, as it correctly reduces to the appropriate two-dimensional result in eq.~(\ref{Poisson2d}). 
Given that Hamiltonian fields involving a two-dimensional symplectic form satisfy $\partial_i A_i=0$ ($i\in\{1,2\}$), all such fields generate area-preserving diffeomorphisms. These will then act independently on a family of two-dimensional subspaces that foliate the four-dimensional spacetime. 

Another type of symplectomorphism one may consider is the case in which the symplectic form coefficients $\Omega_{\mu\nu}$ become complex in Euclidean signature. An example has already been provided above in section~\ref{sec:self-dual}, where eqs.~(\ref{Amukhat}, \ref{khatdef}) define a Hamiltonian vector field with 
\begin{equation}
\Omega_{\mu\nu}=B_i\bar{\eta}^i_{\mu\nu},\quad \vec{B}^2=0.
\label{Omegacomplex}
\end{equation}
In order to satisfy the second condition, the coefficients of $\vec{B}$ must become complex, but we will not need to use the full language of complex symplectic geometry in what follows. We will, however, need the following property for such symplectomorphisms:
\begin{equation}
\Omega_{\mu\alpha}\Omega_{\mu\beta}=0,
\label{Omeganull}
\end{equation}
which follows from the condition $\vec{B}^2=0$ as well as the known property of 't Hooft symbols
\begin{equation}
    \bar{\eta}^i_{\mu\alpha}\bar{\eta}^j_{\mu\beta}=
    \delta^{ij}\delta_{\alpha\beta}+\epsilon^{ijk}\bar{\eta}^k_{\alpha\beta}.
\end{equation}
Once again, we can define a Poisson bracket for these complex symplectomorphisms, and it is given by the final expression in eq.~(\ref{PoissonA}) as before. 

\begin{figure}
  \begin{center}
    \scalebox{0.6}{\includegraphics{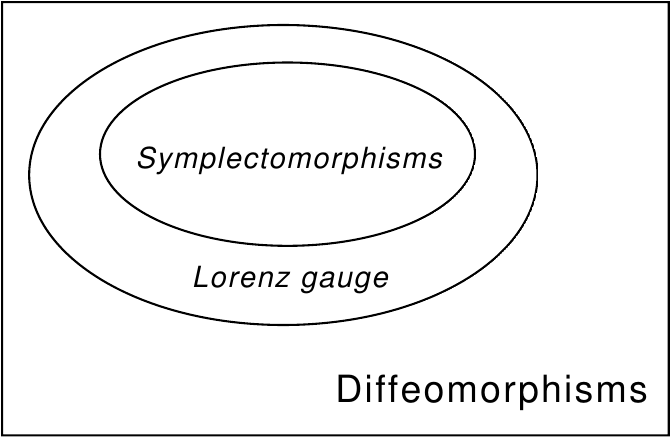}}
    \caption{Schematic view of symplectomorphisms, as a subset of
      volume-preserving diffeomorphisms. The latter are associated
      with abelian gauge fields in the Lorenz gauge.}
    \label{fig:diffplot2}
  \end{center}
\end{figure}

Equation~(\ref{symplectic2}) implies the antisymmetry property
$\Omega_{\mu\nu}=-\Omega_{\nu\mu}$. It then follows from
eq.~(\ref{AOmega}) that any Hamiltonian gauge field satisfies
eq.~(\ref{Lorenz}), and hence is in the Lorenz gauge. However,
symplectomorphisms are a smaller subgroup than mere volume-preserving
diffeomorphisms, thus the class of Hamiltonian vector fields
constitutes a special family of abelian solutions, that restricts to a
subsector of abelian gauge theory, rather than simply being a gauge
choice. Following figure~\ref{fig:diffplot}, we represent this
schematically as shown in figure~\ref{fig:diffplot2}.

Let us now ask what the special set of Hamiltonian gauge fields
corresponds to physically. One example has already been given: the
self-dual abelian fields of eqs.~(\ref{Amukhat}, \ref{khatdef}) are
all Hamiltonian, where the coefficients of the relevant symplectic
form are given by eq.~(\ref{Omegacomplex}). This is clearly not the most general case, however, as one could also have abelian gauge fields based on two- or four-dimensional real symplectomorphisms. We will see an example of the former in what follows.

\subsection{The diffeomorphism algebra of lightcone gauge electromagnetism}
\label{sec:EM}

As a novel application of the ideas of this section, let us elucidate
the diffeomorphism algebra of lightcone electromagnetism. If we restrict to real solutions
of the gauge field in Lorentzian signature, we must analytically
continue eq.~(\ref{Amukhat}) appropriately, and add a complex
conjugate term as follows:
\begin{equation}
  A_\mu=\hat{k}_\mu\phi+\hat{k}_\mu^\dag \phi^\dag.
  \label{AmuEM}
\end{equation}
The two terms consist of a self-dual and anti-self dual contribution
respectively, and thus the diffeomorphism generated by $A_\mu$
consists of a sum of two area-preserving diffeomorphisms in a
self-dual and anti-self-dual null plane respectively. These are known
as $\alpha$- and $\beta$-planes respectively, and we may associate a
$\beta$-plane with a given $\alpha$-plane by demanding that their
respective tangent bivectors are related by complex conjugation in
Lorentzian signature.

Without loss of generality, we may choose a particular lightcone gauge
defined through the lightcone coordinates
\begin{equation}
  u=\frac{t-z}{\sqrt{2}},\quad v=\frac{t+z}{\sqrt{2}},\quad
  X=\frac{x+iy}{\sqrt{2}},\quad Y=\frac{x-iy}{\sqrt{2}},
  \label{uvXY}
\end{equation}
where $(t,x,y,z)$ are Cartesian coordinates in Lorentzian signature,
and the line elements in each coordinate system are given by
\begin{equation}
  ds^2=dt^2-dx^2-dy^2-dz^2=2dudv-2dXdY.
  \label{ds2}
\end{equation}
For the differential operator appearing in eq.~(\ref{AmuEM}), we will
take the explicit form (in the $(u,v,X,Y)$ system)
\begin{equation}
  \hat{k}_\mu=(0,\partial_Y,\partial_u,0)\quad\Rightarrow\quad
  \hat{k}^\mu=(\partial_Y,0,0,-\partial_u),
  \label{kmuform}
\end{equation}
which corresponds to $(B_1,B_2,B_3)=(-i,1,0)$ in eq.~(\ref{Amukhat}),
as shown in ref.~\cite{Armstrong-Williams:2022apo}. The
diffeomorphisms generated by the first term of eq.~(\ref{AmuEM}) are
then area-preserving in the infinite family of $(u,Y)$ planes
given parametrically by
\begin{equation}
  x^\mu=x_0^\mu+\lambda_1(1,0,0,-1)+\lambda_2(0,1,-i,0)
  \label{xmudef1}
\end{equation}
in Cartesian coordinates. Here the first term on the right-hand side
is a constant offset telling us which $\alpha$-plane we are on, and
the remaining two terms contain vectors in the $u$- and $Y$-directions
respectively. The imaginary piece in the final term corresponds to the
well-known fact that null planes with real coordinate values cannot
exist in Lorentzian signature, but would instead be real in (2,2)
signature (in this case corresponding to the replacement $y\rightarrow
iy$). The $\beta$-plane that is conjugate to eq.~(\ref{xmudef1}) is
given by
\begin{equation}
  x^\mu=x_0^\mu+\lambda_1(1,0,0,-1)+\lambda_2(0,1,i,0),
  \label{xmudef2}  
\end{equation}
and corresponds to the differential operator
\begin{equation}
  \hat{k}^{\dag}_\mu=(0,\partial_X,0,\partial_u)\quad\Rightarrow\quad
  \hat{k}^{\dag\mu}=(\partial_X,0,-\partial_u,0)
  \label{kdagmu}
\end{equation}
in the lightcone coordinate system.\footnote{Note that the complex
conjugate relation between $k^\mu$ and $k^{\dag\mu}$ is required to
hold in Lorentzian signature. That eq.~(\ref{kdagmu}) is indeed the
complex conjugate of eq.~(\ref{kmuform}) in the Lorentzian Cartesian
coordinates $(t,x,y,z)$ follows from the fact that complex conjugation
reverses the roles of $X$ and $Y$. Thus, $\partial_Y$ is replaced by
$\partial_X$, and the $X$ and $Y$ coordinates interchanged, in going
from eq.~(\ref{kmuform}) to eq.~(\ref{kdagmu}).} We can see that
eq.~(\ref{kdagmu}) generates diffeomorphisms in the particular $(u,X)$
plane that is related to a given $(u,Y)$ plane by complex conjugation.

\begin{figure}
  \begin{center}
    \scalebox{0.6}{\includegraphics{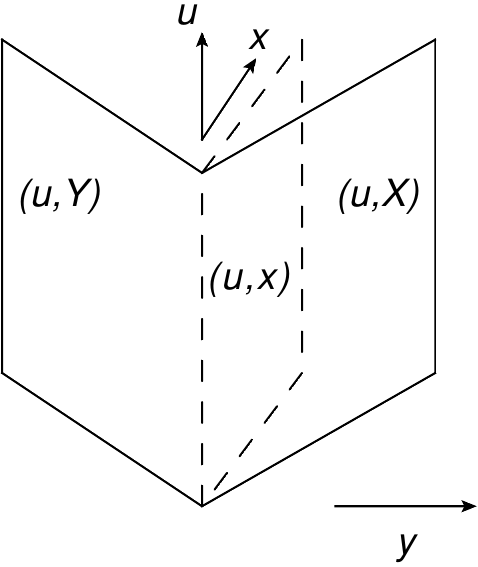}}
    \caption{Each term in eq.~(\ref{AmuEM}) generates area-preserving
      diffeomorphisms in given $(u,Y)$ or $(u,X)$ planes, shown on the
      left and right respectively, where the $y$ components are pure imaginary. Summing these two terms projects
      these diffeomorphisms into the $(u,x)$ plane if $\phi\in{\mathbb
        R}$, and the projection preserves the area-preserving
      property.}
    \label{fig:xufig}
  \end{center}
\end{figure}

We have thus found that the general lightcone gauge field of
eq.~(\ref{AmuEM}), which is neither self-dual nor anti-self-dual,
generates a combination of two area-preserving diffeomorphisms, in
$(u,Y)$ and $(u,X)$ planes respectively. These two diffeomorphisms are
not arbitrary, but linked by the fact that the total gauge field must
be real in Lorentzian signature. We therefore expect the ``kinematic
algebra'' of lightcone gauge electromagnetism to be some subgroup of
the product group
\begin{equation}
  {\rm Diff}_{(u,Y)}\times {\rm Diff}_{(u,X)},
\label{prodgroup}
\end{equation}
where we denote the area-preserving diffeomorphism group associated
with the family of planes $Z$ by ${\rm Diff}_Z$. To find this, we may
substitute eqs.~(\ref{kmuform}, \ref{kdagmu}) into eq.~(\ref{AmuEM})
to get (in lightcone coordinates)
\begin{equation}
  (A^u,A^v,A^X,A^Y)=(\partial_Y\phi+\partial_X\phi^\dag,0,-\partial_u\phi^\dag,
  -\partial_u\phi).
  \label{Amures1}
\end{equation}
In order to understand which diffeomorphisms this generates, it is convenient to transform from $X$ and $Y$ to the Cartesian coordinates $x$ and $y$, keeping the lightcone coordinates $u$ and $v$ as is. The resulting transformed field is then given by
\begin{equation}
  (A^u,A^v,A^x,A^y)=\frac{1}{\sqrt{2}}\left(\partial_x(\phi+\phi^\dag)+i\partial_y(\phi-\phi^\dag),0,
  -\partial_u(\phi+\phi^\dag),-i\partial_u(\phi-\phi^\dag)\right),
  \label{Amures2}
\end{equation}
Eq.~(\ref{Amures2}) generates diffeomorphisms whose integral curves have components
in the $u$ direction, but also the $x$ and $y$
directions. Furthermore, the $x$- and $y$-components are independent,
given that they are governed by the real and imaginary parts of $\phi$
respectively. Thus, a general $A^\mu$ generates diffeomorphisms in the
three-dimensional volume spanned by $(u,x,y)$. In fact, they preserve
volume, given that eq.~(\ref{Amures2}) implies
\begin{equation}
  \partial_u A^u+\partial_x A^x+\partial_y A^y=0.
  \label{volcon}
\end{equation}
In general then, the diffeomorphism algebra of lightcone gauge
electromagnetism is a 3d-volume-preserving subgroup of the product
group of eq.~(\ref{prodgroup}). Interestingly, something more special
happens if the scalar field $\phi$ is itself real ($\phi\in{\mathbb
  R}$). Then eq.~(\ref{Amures1}) simplifies to
\begin{equation}
  (A^u,A^v,A^X,A^Y)=\left((\partial_X+\partial_Y)\phi,0,-\partial_u \phi,-\partial_u\phi
  \right).
  \label{Amures3}
\end{equation}
Again transforming from $(X,Y)$ to $(x,y)$, one finds that the only non-zero components of the gauge field are
\begin{equation}
  A^u=\sqrt{2}\partial_x\phi,\quad A^x=-\sqrt{2}\partial_u\phi.
  \label{Aux}
\end{equation}
We also find
\begin{equation}
  \partial_u A^u+\partial_x A^x=0,
  \label{volcon2}
\end{equation}
so that $A^\mu$ generates area-preserving diffeomorphisms in the
$(u,x)$ plane. There is a pleasing geometric interpretation of this,
as shown in figure~\ref{fig:xufig}. Each individual term in
eq.~(\ref{AmuEM}) generates area-preserving diffeomorphisms in a null
plane, where the $y$ coordinates for points on the plane are pure imaginary, and thus do not show up in real
Lorentzian coordinates. However, summing the two terms in
eq.~(\ref{AmuEM}) for $\phi\in{\mathbb R}$ means that we keep only the
projection into the $(u,x)$ plane. The fact that the resulting
diffeomorphisms are then area-preserving means that this projection
preserves the area-preserving property from the two original planes.
Alternatively, one may consider the case in which the field $\phi$ is pure imaginary:
\begin{equation}
    \phi=i\xi,\quad \xi\in{\mathbb R}.
    \label{phixi}
\end{equation}
Then eq.~(\ref{Amures1}) becomes 
\begin{equation}
    (A^u,A^v,A^X,A^Y)=\Big(i(\partial_Y-\partial_X),0,i\partial_u\xi,
    -i\partial_u\xi\Big)
    \label{Amuycase}
\end{equation}
from which one finds non-zero components after transforming to $(x,y)$
\begin{equation}
A^u=-\sqrt{2}\partial_y\xi,\quad A^y=\sqrt{2}\partial_u\xi,
\label{Auyycase}
\end{equation}
and thus
\begin{equation}
    \partial_u A^u+\partial_y A^y=0.
    \label{areapreserveycase}
\end{equation}
We now have area-preserving diffeomorphisms in the $(u,y)$ plane, where again the area-preserving property is inherited, via a projection, from the original $\alpha$- and $\beta$-planes. To summarise, in both the pure real and imaginary $\phi$ cases, the kinematic algebra is one of real two-dimensional symplectomorphisms, i.e. the closed subgroup of full four-dimensional symplectomorphisms that we discussed in section~\ref{sec:symplect}.

\subsection{Gauge dependence of the diffeomorphism algebra}
\label{sec:gaugedep}

Before moving on to discuss interacting theories, let us also note
that the abelian / linearised context allows us to examine the
issue of gauge-dependence of kinematic algebras. As mentioned in the
introduction, precisely how kinematic algebras of a given theory
depend upon the choice of gauge remains an open question. Indeed, this
has only recently been explored for the best-known kinematic algebra,
namely that of area-preserving diffeomorphisms for self-dual
Yang-Mills theory in the lightcone gauge. As
ref.~\cite{Bonezzi:2023pox} has shown, the kinematic algebra of
self-dual Yang-Mills in other gauges is not expected to be a strict
Lie algebra, but may instead involve a potentially infinite number of
higher brackets in the $\text{BV}_\infty^\Box$ formalism.

\begin{figure}
  \begin{center}
    \scalebox{0.6}{\includegraphics{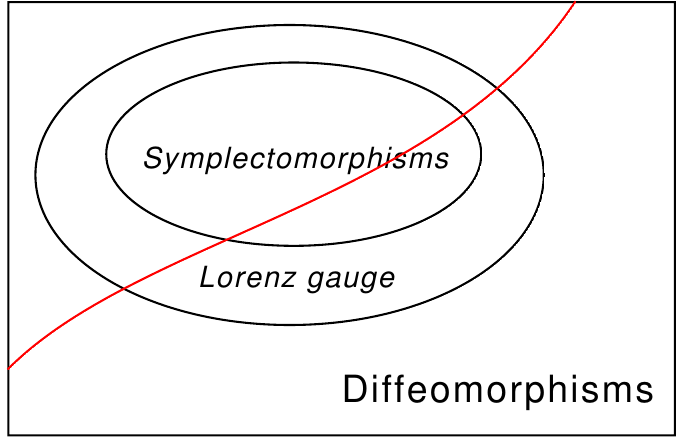}}
    \caption{The set of all physically equivalent abelian gauge fields
      (related by a gauge transformation) shows up as a line -- shown
      in red -- in the space of all possible diffeomorphisms.}
    \label{fig:diffplot3}
  \end{center}
\end{figure}

In an abelian theory, all gauge fields are associated with the Lie
algebra of diffeomorphisms, which does indeed constitute a strict Lie
algebra. However, points in the space of diffeomorphisms, as shown in
figure~\ref{fig:diffplot}(a), constitute particular gauge fields, in a
given gauge. If we instead vary the gauge and consider a so-called
{\it gauge orbit} in the space of gauge fields, this will appear as a
line in the space of diffeomorphisms, as exemplified in
figure~\ref{fig:diffplot3}.

Let us now consider a gauge field that may be chosen to generate a
symplectomorphism. The explicit form of arbitrary gauge fields in the
same orbit is then obtainable as
\begin{equation}
  A_\mu=\hat{k}_\mu \phi+\partial_\mu \chi,
  \label{Amuorbit}
\end{equation}
for some function $\chi$. It is not true in general that
\begin{equation}
  \partial_\mu\chi=\Omega_{\mu\nu}\partial_\nu\chi'
  \label{chi'}
\end{equation}
for some $\chi'$. Thus, a general gauge transformation will lead to a
vector field that does not preserve the symplectic form. If $\chi$ is
harmonic ($\partial^2\chi=0$), then the gauge field will at least
remain in the Lorenz gauge, but this itself will no longer be true if
we consider non-harmonic functions $\chi$.\footnote{In the self-dual sector, the maximal subset of residual symmetries which preserve the light-cone gauge, and hence the Lie kinematic algebra, was studied in \cite{Campiglia:2021srh,Nagy:2022xxs}. This was shown to double copy exactly to the residual subset of symmetries in self-dual gravity which preserve the Plebanski form of the action. More generally, the role of gauge choices and their residual symmetries in extending the double copy beyond flat backgrounds was explored in \cite{Liang:2023zxo}.} Thus, by varying $\chi$, we
will gradually move out of the special subgroups of the diffeomorphism
algebra shown in figure~\ref{fig:diffplot3}. As we will see in the
following section, in certain cases we can build kinematic algebras
for interacting theories by relating these to the diffeomorphism
algebra of an abelian theory. Thus, the gauge-dependence of the
diffeomorphism algebra provides a direct analogue of the
gauge-dependence of non-abelian kinematic algebras.

\section{Interacting theories and kinematic algebras}
\label{sec:interactions}

In the previous section, we have seen that we can classify certain
meaningful subgroups of diffeomorphism algebras of abelian gauge
fields, which have a definite physical interpretation. As we have
stressed repeatedly throughout, however, this is not what is usually
meant by the kinematic algebra of a gauge theory, which is instead
associated with interaction terms. In this section, we argue that the ideas of the
previous section remain useful, in allowing us to look for interacting
theories that have straightforward kinematic algebras, and our starting point will be
to reinterpret the well-known case of self-dual Yang-Mills theory.

\subsection{Self-dual Yang-Mills theory and the Poisson bracket}
\label{sec:SDYM}

As first presented in ref.~\cite{Monteiro:2014cda}, and discussed also
above, we can obtain the field equation for self-dual Yang-Mills
theory in lightcone gauge by making the ansatz of
eq.~(\ref{Amuakhat}), where $\hat{k}_\mu$ satisfies the conditions of
eq.~(\ref{khat}). Substituting this into the Yang-Mills equations
yields
\begin{equation}
  \hat{k}_\nu\left[\partial^2\phi^a+2gf^{abc}(\hat{k}^\mu\phi^b)
    (\partial_\mu\phi^c)\right]=0.
  \label{SDYM}
\end{equation}
Introducing the matrix-valued scalar field
\begin{equation}
  {\bf \Phi}=\phi^a{\bf T}^a,
  \label{Phidef}
\end{equation}
eq.~(\ref{SDYM}) assumes the form
\begin{equation}
  \partial^2 {\bf \Phi}+2ig\left\{\left[{\bf \Phi},{\bf \Phi}\right]\right\}=0,
  \label{SDYM2}
\end{equation}
where we have introduced the double bracket
\begin{equation}
  \left\{\left[{\bf \Phi},{\bf \Phi}\right]\right\}=i
  f^{abc}\Omega^{\mu\nu}(\partial_\mu\phi^b)(\partial_\nu\phi^c){\bf T}^a.
  \label{bracket}
\end{equation}
Comparison with eqs.~(\ref{Lie}, \ref{PoissonA}) reveals that this
bracket combines a Poisson bracket in the kinematic variables, with
the conventional Lie bracket of two colour generators. This in turn
means that the interaction term of the theory carries structure
constants of both the (colour) Lie algebra, and the kinematic Poisson
algebra. Thus, these structure constants will appear alongside each
other in all perturbative solutions of the theory, such that BCJ
duality is manifest. 

There is an interesting way to reinterpret this result, based on the
ideas of the previous section. In particular, if we consider an
abelian (or linearised non-abelian) gauge theory, we can focus on the
subsector of the theory (in a particular gauge), such that the gauge
field is Hamiltonian, and defined according to a differential operator
$\hat{k}_\mu$ satisfying the conditions of eq.~(\ref{khat}). We can then consider extending the theory to make it interacting, by using the building blocks that already exist at linear level. That is, we can make a Poisson bracket out of the symplectic form coefficients that already exist in the Hamiltonian gauge field, and then combine this with the colour Lie bracket to make an interaction term. There is then a sense in which the kinematic algebra of
the interacting theory is inherited from the diffeomorphism algebra
that already appears at linearised level. This link is made more
precise by eq.~(\ref{phi3def}), which expresses the fact that the Lie
bracket of two Hamiltonian fields is itself Hamiltonian, but where the
Hamiltonian of the resulting field is given by the Poisson bracket of
the two original Hamiltonians. Thus, the Poisson bracket in the scalar
formulation of the theory encodes the underlying Lie algebra of the
Hamiltonian vector fields $A_\mu^a$. 

This suggests a general recipe for constructing
non-linear extensions of linearised gauge theory, where the
interaction term is characterised by a combined Poisson / Lie bracket. By choosing different symplectic form coefficients $\Omega_{\mu\nu}$, one may obtain different interacting theories. One may also replace $\Omega_{\mu\nu}$ with a more general antisymmetric matrix, which does not satisfy the symplectic form conditions, but nevertheless yields a closed algebra of abelian gauge fields. Interestingly, eq.~(\ref{SDYM2}), with a
general antisymmetric matrix $\Omega_{\mu\nu}$ entering the Poisson
bracket, was considered in e.g. refs.~\cite{BjerrumBohr:2012mg,Chacon:2020fmr} as a way of 
generalising (anti-)self-dual kinematic algebras. Here we provide
a more systematic basis for this equation, and we are also able to
obtain a direct geometric interpretation of the algebra: it
corresponds to the diffeomorphism subgroup
associated with the linearised gauge fields $A_\mu^a$. Unlike the
(anti-)self-dual cases, diffeomorphisms associated with a general antisymmetric $\Omega_{\mu\nu}$ will not be
area-preserving in general. To see this, note that a general antisymmetric matrix in four
spacetime dimensions has rank $\leq 4$. Thus, the resulting
diffeomorphisms, whilst still being volume-preserving, are not
guaranteed to reduce to acting in lower-dimensional
hypersurfaces.

\subsection{Electromagnetism coupled to scalar matter}
\label{sec:EMmat}

In the previous section, we have seen that one may construct
non-linear theories whose kinematic algebras are based on the
diffeomorphism algebra that already exists at linearised level. Those cases involved constructing a double bracket consisting of a colour Lie bracket, combined with a kinematic (Poisson) bracket. This is an intrinsically non-abelian construction, and our aim in this section is to show that a similar idea can be used, even if the gauge field is abelian. Let us start with an abelian gauge field $A_\mu$ and, in line with the examples of the previous section, we will restrict to the subset of Hamiltonian fields, such that 
\begin{equation}
    A_\mu=\Omega_{\mu\nu}\partial^\nu\phi,
    \label{AmuHamiltonian}
\end{equation}
for some scalar field $\phi$. The general vacuum field equation for $A_\mu$ is
\begin{equation}
    \partial^2 A_\mu-\partial_\mu(\partial\cdot A)=0
    \label{EM1}
\end{equation}
which, in the case of Hamiltonian vector fields, yields
\begin{equation}
    \partial^2 \phi=0.
    \label{EM2}
\end{equation}
As before, considering Hamiltonian vector fields implies that there is a symplectic form, which can in turn be used to build a Poisson bracket. We can then look for non-linear extensions of eq.~(\ref{EM2}), and investigate whether any of them can be seen as truncations of some more complete theory. If we wish to preserve the fact that $A_\mu$ is abelian, then there is no non-zero Poisson bracket involving $\phi$ with itself:
\begin{equation}
\{\phi,\phi\}=0.
\label{phiphi0}
\end{equation}
Instead, we can consider an additional scalar field $\psi$, which at linear level satisfies the Klein-Gordon equation
\begin{equation}
    \partial^2\psi=0.
    \label{psieq}
\end{equation}
This equation can then be extended non-linearly as 
\begin{equation}
    \partial^2\psi+c_1\{\psi,\phi\}=0,
    \label{phipsieq}
\end{equation}
where we have used the fact that one may make a Poisson bracket out of the scalar field $\psi$, and the scalar that enters the gauge field via eq.~(\ref{AmuHamiltonian}). The question then naturally arises as to whether eq.~(\ref{phipsieq}) is a physically consistent 
theory, which may in turn depend on the value of the coefficient $c_1$. Indeed, in order to be talking about a gauge field interacting with $\psi$ at all, it must be the case that eq.~(\ref{phipsieq}) -- or some generalisation of it -- be gauge-covariant. Let us consider gauge transformations that preserve the Hamiltonian nature of $A_\mu$:
\begin{equation}
    A_\mu\rightarrow A'_\mu = A_\mu+\partial_\mu\chi,
    \label{Amutrans}
\end{equation}
where the corresponding gauge transformation for the scalar field is
\begin{equation}
\psi\rightarrow\psi'=e^{-ie\chi}\psi,
\label{psitrans}
\end{equation}  
and where there will be a restriction on $\chi$:
\begin{equation}
\partial_\mu\chi=\Omega_{\mu\nu}\partial^\nu \alpha
\label{chicond}
\end{equation}
for some $\alpha$. Using eq.~(\ref{AmuHamiltonian}), one may rewrite eq.~(\ref{phipsieq}) as
\begin{equation}
    \partial^2\psi+c_1 A_\mu \partial^\mu\psi=0,
    \label{phipsieq2}
\end{equation}
which under a gauge transformation satisfies
\begin{equation}
    \partial^2 \psi'+A'_\mu \partial^\mu\psi'=0\quad\rightarrow\quad
    \partial^2 \psi+A_\mu \partial^\mu\psi+\Delta=0,
    \label{gaugetranseq}
\end{equation}
with
\begin{equation}
\Delta=(c_1 - 2ie)(\partial_\mu\chi)(\partial^\mu \psi)
-ie c_1A_\mu (\partial^\mu\chi)\psi-(iec_1+e^2)(\partial_\mu\chi)
(\partial^\mu\chi)\psi.
\label{Deltadef1}
\end{equation}
There is no solution for $c_1$ that yields $\Delta=0$, corresponding to the well-known fact that one must add a seagull vertex to scalar QED in order to make it gauge-invariant. Let us then correct eq.~(\ref{phipsieq2}) to read
\begin{equation}
    \partial^2\psi+c_1 A_\mu\partial^\mu\psi+c_2A^\mu A_\mu\psi=0.
    \label{phipsieq3}
\end{equation}
Upon doing so and carrying through the above steps, the difference between field equations in different gauges of eq.~(\ref{Deltadef1}) becomes instead 
\begin{equation}
\Delta=(c_1 - 2ie)(\partial_\mu\chi)(\partial^\mu \psi)
+(2c_2-ie c_1)A_\mu (\partial^\mu\chi)\psi+(c_2-iec_1-e^2)(\partial_\mu\chi)
(\partial^\mu\chi)\psi.
\label{Deltadef2}
\end{equation}
The unique solution for $\Delta=0$ is $(c_1,c_2)=(2ie,-e^2)$, so that the gauge-invariant scalar field equation is
\begin{equation}
\partial^2\psi+2ie \{\psi,\phi\}-e^2 A^\mu A_\mu\psi=0.
    \label{phipsieq4}
\end{equation}
This has a cubic term, arising from a quartic interaction in the Lagrangian for the theory that gives rise to this equation of motion. It is not then true in general that there is a straightforward kinematic Lie algebra, i.e. such that there are up-to-quadratic terms in the field equation only.

We can of course find the subsector of solutions of eq.~(\ref{phipsieq4}) for which the cubic term vanishes, and the criterion for this is straightforward. From eq.~(\ref{AmuHamiltonian}), the final term in eq.~(\ref{phipsieq4}) will vanish provided
\begin{equation}\label{OmegaSquare0}
    \Omega_{\mu\alpha}{\Omega^\mu}_\beta=0.
\end{equation}
This property is satisfied by self-dual field configurations in the light-cone gauge. In this case the gauge field must satisfy eq.~\eqref{khat}, which for a Hamiltonian vector field written in terms of a symplectic form, amounts to eq.~\eqref{OmegaSquare0}. As discussed previously in section~\ref{sec:self-dual}, linearised self-dual solutions can be written in terms of a superposition of 't Hooft symbols, which in turn allows for a simple geometric interpretation of the generated diffeomorphisms. Thus, by the general arguments of section~\ref{sec:self-dual}, such fields will generate area-preserving diffeomorphisms in either $\alpha-$ or $\beta-$planes. 

Let us therefore consider the case of abelian self-dual solutions in eq.~\eqref{phipsieq4}, such that it straightforwardly reduces to 
\begin{equation}
    \partial^2\psi+2ie\{\psi,\phi\}=0.
    \label{SDEMpsi}
\end{equation}
We now wish to ask whether this equation can be obtained from a top-down approach, such that it corresponds to an equation of motion in a particular theory. As we have seen that in attempting to extend electromagnetism with a non-trivial Poisson bracket we arrive at a self-dual photon coupled to a scalar, let us consider the abelian gauge field to be self-dual. To this end, consider the following Lagrangian
\begin{equation}\label{LBF}
    \mathcal{L} = B_{\mu\nu}F_{-}^{\mu\nu} +(D^\mu \psi)^\dag (D_\mu\psi), 
\end{equation}
where we have introduced the covariant derivative
\begin{equation}
  D_\mu=\partial_\mu+ieA_\mu.
  \label{Dmudef}
\end{equation}
Here $B_{\mu\nu}$ are the components of an anti-self dual two-form and $F_{-}$ is the anti-self-dual part of the field strength. The field $B_{\mu\nu}$ acts as a Lagrange multiplier in the action, enforcing the self-duality of the abelian field strength. The theory thus corresponds to self-dual electromagnetism coupled to a complex scalar field. 

\begin{figure}
    \begin{center}
        \scalebox{0.7}{\includegraphics{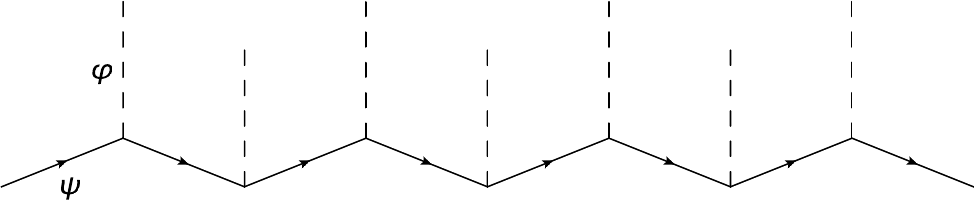}}
        \caption{Example Feynman diagram generated by the theory of eq.~(\ref{LBFscal}).}
        \label{fig:ladder}
    \end{center}
\end{figure}

By adopting a light-cone gauge and integrating out two of the three independent components of $B_{\mu\nu}$, we enforce a Hamiltonian form for the gauge field as well as the condition in eq.~\eqref{OmegaSquare0} (see e.g. ref.~\cite{Monteiro:2022nqt} for an example of this procedure in the non-abelian case). The result is an action only in terms of scalar degrees of freedom
\begin{equation}
    \mathcal{L} = \bar{\phi}\,\partial^2\phi - \psi^{\dagger}\partial^2\psi - i\Omega^{\mu\nu}\partial_{\nu}\phi\left(\psi^{\dagger}\partial_{\mu}\psi  -\psi\partial_{\mu}\psi^{\dagger}\right),
    \label{LBFscal}
\end{equation}
where $\bar{\phi}$ is the final component of $B_{\mu\nu}$ and the quartic interaction is not present as a consequence of eq.~\eqref{OmegaSquare0}. We can recognise the first term as a linearised form of the self-dual Yang-Mills action in light-cone gauge~\cite{Chalmers:1996rq}, where $\phi$ and $\bar{\phi}$ are interpreted as the positive and negative helicity degrees of freedom of the gauge field respectively. By integrating by parts and making use of eqs.~\eqref{Poisson2d},~\eqref{Poissonrel} we obtain
\begin{equation}\label{ScalarAction}
    \mathcal{L} = \bar{\phi}\,\partial^2\phi - \psi^{\dagger}\partial^2\psi + 2ie\phi\{\psi,\psi^{\dagger}\}.
\end{equation}
Thus, we see that the symplectic form present in the linearised self-dual gauge field induces a Poisson bracket structure in the interaction vertex. The equations of motion for this theory are
\begin{align}
    \partial^2\phi &= 0, \label{eom1}\\
    \partial^2\bar{\phi} + 2ie \{\psi,\psi^{\dagger}\} &=0, \label{eom2} \\
    \partial^2\psi + 2ie \{\psi,\phi\} &=0, \label{eom3} \\
    \partial^2\psi^{\dagger} - 2ie \{\psi^{\dagger},\phi\} &=0. \label{eom4}
\end{align}
Equation~\eqref{eom3} corresponds precisely to eq.~\eqref{SDEMpsi}. In all cases, the non-linear terms inherit a Poisson bracket structure from the symplectic form present in the gauge field. This theory can be used to generate tree-level amplitudes in which a backbone of scalar $\psi$ exchanges radiates a series of photon states $\phi$, as exemplified in figure~\ref{fig:ladder}. This is clearly only a subset of the amplitudes contained in the full theory of electromagnetism coupled to a scalar field. However, it is interesting that a subset indeed exists, where the vertices can be associated with kinematic structure constants inherited from a Poisson bracket. 

As in the Yang-Mills case, the kinematic algebra ceases to be straightforward once both self-dual and anti-self-dual degrees of freedom are included. To see this explicitly, let us view the action of eq.~\eqref{LBF} as a sector of full scalar QED, where the action for this theory is
\begin{equation}
   {\cal L}=-\frac14 F^{\mu\nu}F_{\mu\nu}+(D^\mu \psi)^\dag (D_\mu\psi),
   \label{SscalarQED}
\end{equation}
We now follow a standard procedure for reducing this action to one only in terms of the propagating degrees of freedom, as was done for Yang-Mills theory in ref.~\cite{Chalmers:1998jb}. We choose the light-cone gauge $A_u = 0$. Then, in the scalar QED action, $A_v$ appears quadratically and can be functionally integrated out. Labelling the remaining components of the gauge field as $A_Y=A, A_X=\bar{A}$, we obtain an action 
\begin{align}\label{SscalarQEDcs}
    \mathcal{L} &= -\bar{A}\partial^2A -\psi^{\dag}\partial^2\psi 
    + ieA\left[\psi^{\dag}\partial_X\psi - \psi\partial_X\psi^{\dag}-\frac{\partial_X}{\partial_u}\left(\psi^{\dag}\partial_u\psi - \psi\partial_u\psi^{\dag}\right)\right] \nonumber\\
    &+ ie\bar{A}\left[\psi^{\dag}\partial_{Y}\psi - \psi\partial_{Y}\psi^{\dag}-\frac{\partial_{Y}}{\partial_u}\left(\psi^{\dag}\partial_u\psi - \psi\partial_u\psi^{\dag}\right)\right]
    -2e^2A\bar{A}|\psi|^2
    +\frac{e^2}{2}\left[\frac{1}{\partial_u}\left(\psi^{\dag}\partial_u\psi - \psi\partial_u\psi^{\dag}\right)\right]^2
\end{align}
The fields $A$ and $\bar{A}$ are interpreted as the positive and negative helicity degrees of freedom of the gauge field. The action contains three-point $(A\,\psi\,\psi^\dag)$ and $(\bar{A}\,\psi\,\psi^\dag)$ vertices, a four-point $(A\,\bar{A}\,\psi\,\psi^\dag)$ vertex, and a quartic scalar vertex. This theory will generate the amplitudes exemplified by figure~\ref{fig:ladder}, and more besides. However, the quartic vertices in eq.~(\ref{SscalarQEDcs}) will not enter the ladder amplitudes described above, which result upon keeping only the $(A\,\psi\,\psi^\dag)$ vertex. To make contact with eq.~(\ref{LBFscal}) explicitly, we may perform the field redefinitions $A=\partial_u\phi, \bar{A}=\partial_u^{-1}\bar{\phi}$, such that keeping only the first cubic vertex in eq.~(\ref{SscalarQEDcs}) and disregarding the quartic vertices yields 
\begin{equation}
    \mathcal{L} = \bar{\phi}\partial^2\phi -\psi^{\dag}\partial^2\psi -ie\phi\left[\partial_u\left(\psi^{\dag}\partial_X\psi - \psi\partial_X\psi^{\dag}\right)-\partial_X\left(\psi^{\dag}\partial_u\psi - \psi\partial_u\psi^{\dag}\right)\right],
\end{equation}
where we have integrated by parts. After a little algebra, the vertex structure reduces to a Poisson bracket and we obtain eq.~\eqref{ScalarAction}.

% One way to do this is to extend the Lagrangian of eq.~(\ref{LBF}) to include a quadratic term in the auxiliary field:
% \begin{equation}
%   {\cal L}=B_{\mu\nu}F_{-}^{\mu\nu}+\frac18 B_{\mu\nu}B^{\mu\nu}
%   +(D^\mu \psi)^\dag (D_\mu\psi).
%   \label{LBF2}
% \end{equation}
% The field equation for $B_{\mu\nu}$ now sets this to be proportional to the field strength $F_{\mu\nu}$, leading to the usual action for electromagnetism coupled to a scalar. In the light-cone gauge, one can still reduce this action to one in terms of only the propagating scalar degrees of freedom of the gauge field, along with the complex scalar.
Returning to eq.~\eqref{SscalarQEDcs}, we see that we lose any clear sign of the diffeomorphism algebra present in the linearised theory due to the non-vanishing of the quartic interaction. Similar to full Yang-Mills theory, it is the presence of this higher-order interaction that mixes the self-dual and anti-self dual degrees of freedom of the gauge field, and disrupts the possibility of identifying the kinematic algebra with a straightforward Lie algebra. This complication disappears in the self-dual sector, which constitutes a particular truncation of the theory.

In both the self-dual Yang-Mills and scalar QED examples, relevant equations of motion are ``simple'' in that their kinematic algebras terminate at cubic order,
due to a judicious choice of subsector of the full theory. In both
cases, this involves choosing Hamiltonian gauge fields, such that a
Poisson bracket may be used to construct the interaction terms. It is
worth asking whether one may instead use the Lie bracket of gauge
fields, and to consider actions that are manifestly written in terms
of vector gauge fields. That this will not lead to a physical
theory in general is well-documented in the
literature~\cite{Fu:2016plh}. However, there are indeed special cases where this occurs, which have in fact already appeared in the literature. Let us take each case in turn.

\subsection{Non-abelian Chern-Simons theory}
\label{sec:CS}

Non-abelian Chern-Simons theory is a certain gauge theory in three spacetime dimensions, whose topologically non-trivial solutions have led to a variety of applications in mathematical physics, including connections to knot theory (see e.g. ref.~\cite{Baez:1995sj} for a review). In ref.~\cite{Ben-Shahar:2021zww}, the kinematic algebra of this theory was shown to be a simple Lie algebra of volume-preserving diffeomorphisms, if a particular gauge (the Lorenz gauge) was used for the field $A_\mu^a$. Here, we show that this conclusion naturally arises from the ideas of this paper, thus providing an alternative point of view on this result.

We start by considering the following action for abelian Chern-Simons theory:
\begin{equation}
S_{\rm CS}=\frac{k}{4\pi} \int d^3 x\, \frac12\epsilon^{\mu\nu\rho}
A_\mu\partial_\nu A_\rho,
\label{SCS}
\end{equation}
where $k$ is a constant parameter. The field equation for $A_\mu$ is 
\begin{equation}
    \epsilon^{\mu\nu\rho}\left[\partial_\nu A_\rho -\partial_\rho
    A_\nu\right]\equiv \epsilon^{\mu\nu\rho}F_{\nu\rho}= 0,
    \label{CSabel}
\end{equation}
where we have recognised the abelian field strength $F_{\mu\nu}$. This in turn implies $F_{\mu\nu}=0$, such that solutions of the theory are pure gauge. 

Similar to the previous sections, we may regard $A_\mu$ as generating diffeomorphisms, and then look to extend the theory by using the Lie bracket of diffeomorphisms in forming an interaction term. To do this, we can extend $A_\mu$ to make a non-abelian gauge field with components $A_\mu^a$. We may then consider the double bracket
\begin{equation}
    \left[\left[{\bf A},{\bf A}\right]\right]=
    {\bf T}^a\, f^{abc}\left[A^{\mu\,b}\partial_\mu,A^{\nu\,c}\partial_\nu\right],
    \label{Liedouble1}
\end{equation}
written in terms of the gauge field, contracted with colour and kinematic generators:
\begin{equation}
{\bf A}=A^{\mu\,a} {\bf T}^a\partial_\mu.
\label{Afielddef}
\end{equation}
Equation~(\ref{Liedouble1}) consists of a simultaneous colour Lie bracket, and kinematic Lie bracket (i.e. the latter corresponds to a commutator of diffeomorphisms). This is analogous to eq.~(\ref{SDYM2}), whose double bracket contains a colour Lie bracket and a Poisson (kinematic) bracket. For dimensional reasons, we cannot simply add this bracket to the non-abelian version of eq.~(\ref{CSabel}). However, in line with ref.~\cite{Ben-Shahar:2021zww}, we may instead contract eq.~(\ref{CSabel}) with the combination $\epsilon_{\sigma\mu\alpha}\partial^\alpha$ to obtain
\begin{equation}
    \partial^2 A_\sigma-\partial_\sigma(\partial\cdot A)=0,
    \label{CSEOM2b}
\end{equation}
such that a suitable non-abelian generalisation is
\begin{equation}
\partial^2 A^a_\sigma - \partial_\sigma(\partial\cdot A^a)
+\gamma f^{abc}\left[A^b\cdot\partial A_\sigma^c-A^c\cdot\partial
A^b_\sigma\right]=0,
\label{CSnonabel}
\end{equation}
where we have substituted the explicit form of the Lie bracket of two vector fields. In order for this to be a consistent (sub-)theory, the equation of motion must be gauge-covariant, or at least correspond to some suitable gauge-fixing. This will fix the undetermined parameter $\gamma$. To this end, one may use the product rule, and rearrange terms, to show that eq.~(\ref{CSnonabel}) is equivalent to
\begin{equation}
    \partial^\rho
    F_{\rho\sigma}^a+(2\gamma+1)f^{abc}A^b\cdot\partial A_\sigma^c
    +f^{abc}(\partial \cdot A^b)A_\sigma^c
    =0,
    \label{CSnonabel2}
\end{equation}
where
\begin{equation}
    F_{\rho\sigma}^a=\partial_\rho A_\sigma^a-\partial_\sigma
    A_\rho^a- f^{abc} A_\rho^b A_\sigma^c
    \label{Fnonabel}
\end{equation}
denotes a component of the non-abelian field strength. We can then find a  suitably gauge-fixed field equation by setting
\begin{equation}
    \gamma=-\frac12,\quad \partial\cdot A^a=0,
    \label{gammaval}
\end{equation}
after which eq.~(\ref{Fnonabel}) reduces to
\begin{equation}
    \partial^\rho F_{\rho\sigma}^a=0.
    \label{CSeq}
\end{equation}
Given we have introduced an extra derivative above, we can then infer the field equation
\begin{equation}
    F_{\rho\sigma}^a=0,
    \label{CSeq2}
\end{equation}
which eq.~(\ref{gammaval}) tells us is in Lorenz gauge. This is the known field equation of Chern-Simons theory, and substituting eq.~(\ref{gammaval}) into eq.~(\ref{CSnonabel}) reveals that this can be written as
\begin{equation}
\partial^2 A^a_\sigma 
-\frac12 f^{abc}\left[A^b\cdot\partial A_\sigma^c-A^c\cdot\partial
A^b_\sigma\right]=0
\label{CSnonabel3}
\end{equation}
or, when contracted with generators,
\begin{equation}
    \partial^2 {\bf A}-\frac12\left[\left[ {\bf A},{\bf A} \right]\right]=0.
    \label{CSEOM3}
\end{equation}
To see that this agrees with a standard derivation of the non-abelian Chern-Simons equation, note that the action for this theory can be written in components as\footnote{It is more common to introduce the one-form $A\equiv A_\mu^a dx^\mu {\bf T}^a$, where $\{{\bf T}^a\}$ are the generators of the gauge group, and to consider the action $S=\frac{k}{4\pi}\int d^3 x\,{\rm Tr}\left(A\wedge dA+\frac{2i}{3}A\wedge A\wedge A\right)$. This reduces to eq.~(\ref{SCS}) after substituting components.}
\begin{equation}
S_{\rm CS, non-abel.}=\frac{k}{4\pi}\int d^3 x\, \epsilon^{\mu\nu\rho}\left(
\frac12 A_\mu^a\partial_\nu A_\rho^a-\frac16 f^{abc} A_\mu^a A_\nu^b
A_\rho^c\right).
\label{SCS2}
\end{equation}
The field equation for $A_\mu^a$ is then
\begin{equation}
    \epsilon^{\mu\nu\rho}\left[\partial_\nu A_\rho^a-\frac12 f^{abc}
    A_\nu^b A_\rho^c\right]=0.
    \label{CSEOM}
\end{equation}
To compare with our above results, we must contract eq.~(\ref{CSEOM}) with $\epsilon_{\sigma\mu\alpha}\partial^\alpha$ to obtain
\begin{equation}
    \partial^2 A_\sigma^a-\partial_\sigma(\partial\cdot A^a)
    -\frac12 f^{abc}\left[A_\sigma^c\partial\cdot A^b+A^b\cdot\partial A_\sigma^c
    -(b\leftrightarrow c)\right]=0.
    \label{CSEOM2}
\end{equation}
Then choosing the Lorenz gauge $\partial\cdot A^a=0$ yields eq.~(\ref{CSnonabel3}) as required. Our construction of this theory using the double bracket above makes manifest that there is a Lie kinematic algebra. From the top-down point of view, however, we may see why a straightforward Lie kinematic algebra is not manifest at the level of the equation of motion if we are not in Lorenz gauge. Let us return to the full field equation of eq.~(\ref{CSEOM2}), and define an alternative bracket whose vector components are
\begin{equation}
\left[A_1,A_2\right]^\mu_{\rm CS}=a_1\left((\partial\cdot A_1)A_2^\mu-
(\partial\cdot A_2)A_1^\mu\right)+a_2\left(
A_1\cdot\partial A_2^\mu-A_2\cdot\partial A_1^\mu\right),
\label{CSbracket}
\end{equation}
where $\{A_i\}$ are vector fields, 
and $\{a_i\}$ constant parameters. This bracket is skew-symmetric in its arguments, and reduces to the standard Lie bracket for $(a_1,a_2)=(0,1)$. Furthermore, upon choosing the special case $(a_1,a_2)=(1,1)$, we can express eq.~(\ref{CSEOM2}) as
\begin{equation}
\partial^2 A^{a\sigma}-\partial^\sigma (\partial\cdot A^a)-\frac12f^{abc}[A^b,A^c]^\sigma_{\rm CS}=0.
\label{CSbracketEOM}
\end{equation}
As in our previous examples, this contains a double bracket, this time consisting of a Lie bracket in the colour group, and the generalised kinematic bracket of eq.~(\ref{CSbracket}). What prevents the identification of a straightforward kinematic algebra, however, is the fact that the bracket of eq.~(\ref{CSbracket}) does not satisfy the Jacobi identity. Denoting a momentum-space gauge field by $A(p)$, an explicit calculation reveals that
\begin{align}
&\Big[\left[A_1(p_1),A_2(p_2)\right],A_3(p_3)\Big]^\mu_{\rm CS}+
\Big[\left[A_2(p_1),A_3(p_2)\right],A_1(p_3)\Big]^\mu_{\rm CS}+
\Big[\left[A_3(p_1),A_1(p_2)\right],A_2(p_3)\Big]^\mu_{\rm CS}=
\notag\\
\quad & -a_1\left\{
\Big[(p_1\cdot A_1)(p_1\cdot A_2)-(p_2\cdot A_2)(p_2\cdot A_1)\Big]A_3^\mu
+\Big[(p_2\cdot A_2)(p_2\cdot A_3)-(p_3\cdot A_3)(p_3\cdot A_2)\Big]A_1^\mu\right.\notag\\
&\left.+\Big[(p_3\cdot A_3)(p_3\cdot A_1)-(p_1\cdot A_1)(p_1\cdot A_3)\Big]A_2^\mu
\right\}.
\label{Jacobi}
\end{align}
The non-exact nature of the Jacobi identity can be directly traced to the coefficient $a_1$ appearing in eq.~(\ref{CSbracket}), and thus to the additional contribution that supplements the strict Lie bracket of vector fields. This contribution vanishes only in Lorenz gauge in general, such that we indeed see that the kinematic algebra must be a more complicated mathematical structure than a Lie algebra if we go to arbitrary gauges. Similar considerations were applied to four-dimensional Yang-Mills theory in ref.~\cite{Fu:2016plh}, which proposed certain generalisations of Lie algebras as underlying kinematic algebras. An extended discussion of Chern-Simons theory has been given in the ${\rm BV}^\Box_{\infty}$ approach in ref.~\cite{Bonezzi:2022bse}. There, rather than contracting eq.~(\ref{CSEOM}) with $\epsilon_{\sigma\mu\alpha}\partial^\alpha$, a more general procedure is given for extracting a kinematic bracket of gauge fields, based on the ${\rm BV}^\Box_\infty$ algebra underlying the theory. This then turns out to yield a simple Lie bracket after all. 

Next, we examine a second case in which the Lie bracket of vector fields appears in an interacting theory, and which includes the theory of this section as a special case.

\subsection{Semi-abelian Yang-Mills theory}
\label{sec:semiabel}

Recently, ref.~\cite{Edison:2023ulf} introduced an interesting field theory, aimed at unifying diverse examples of theories obeying colour-kinematics duality, as well as developing systematic procedures for constructing BCJ-dual kinematic numerators in scattering amplitudes. The authors refer to this as {\it semi-abelian Yang-Mills theory}, and it has the following Lagrangian:
\begin{equation}
{\cal L}^{\rm semi-YM}=-\frac12{\rm Tr}\left[\bar{\bf F}_{\mu\nu}
{\bf F}^{\mu\nu}\right].
\label{semiabel}
\end{equation}
Here ${\bf F}_{\mu\nu}$ is the field strength for a non-abelian gauge field ${\bf A}_\mu$ valued in the Lie algebra of U($N$). The additional field strength $\bar{\bf F}_{\mu\nu}$ corresponds to a field $\bar{\bf A}_\mu$ associated with the gauge group U(1)$^{N^2}$. In components, both gauge fields will carry a ``colour index'' $a$ taking values in the range $\{1,\ldots N^2\}$, and the Lagrangian takes the form
\begin{equation}
    {\cal L}^{\rm semi-YM}=\bar{A}^{\nu\,a}
    \left[\partial^2 A_\nu^a-\partial_\nu(\partial\cdot A^a)
    -f^{abc}(\partial\cdot A^b)A_\nu^c+\frac12 f^{abc}\left(
    A^b\cdot\partial A_\nu^c-A^c\cdot\partial A_\nu^b
    \right)
    \right].
    \label{semiabel2}
\end{equation}
Upon constructing the Euler-Lagrange equation for $\bar{A}_\mu^a$, one straightforwardly obtains that the field ${\bf A}\equiv A^{\mu\,a}{\bf T}^a\partial_\mu$ satisfies eq.~(\ref{CSEOM3}) in the Lorenz gauge. There is thus again a double-bracket combining the colour Lie bracket with the Lie algebra of diffeomorphisms. Furthermore, the latter are volume-preserving, given the Lorenz gauge condition. As argued in ref.~\cite{Edison:2023ulf}, both self-dual Yang-Mills theory and Chern-Simons theory can be seen as special cases of semi-abelian Yang-Mills theory. From the perspective of this paper, we can perhaps regard semi-abelian Yang-Mills theory as the theory one arrives at upon starting with linearised Yang-Mills theory, and demanding a well-defined kinematic algebra by looking for a double bracket based on the Lie algebra of diffeomorphisms. It would then be interesting to know if this conclusion is unique. We provide a final novel example of an interacting theory containing a similar double bracket in the following section.

\subsection{Fluid mechanics and kinematic algebras}
\label{sec:fluids}

Since its original incarnation involving gauge and gravity theories,
the study of the double copy and its related kinematic algebras has
considerably broadened. Useful for our purposes is
ref.~\cite{Cheung:2020djz}, which considered a non-abelian
generalisation of the Navier-Stokes equation of fluid mechanics:
\begin{equation}
  (\partial_0-\nu\nabla^2)u_i^a+f^{abc}u_j^b\partial_j u_i^c=J_i^a.
\label{NNSE}
\end{equation}
The quantity $u_i^a$ ($i\in\{1,2,3\}$) is the velocity field of a
fluid of viscosity $\nu$, and satisfies the solenoidal requirement $\partial_i u_i^a=0$. The velocity also carries an adjoint index
$a$ associated with a non-abelian colour group, with structure
constants $f^{abc}$. Finally, there is a source current $J_i^a$ on the
right-hand side of eq.~(\ref{NNSE}). This theory was used in
ref.~\cite{Cheung:2020djz} for various purposes, including elucidating
infrared properties of its scattering amplitudes, examining its
kinematic algebra, and exploring its double copy to a bifluid theory,
whose velocity field $u_{i\bar{i}}$ carries two independent spatial
indices. Here we draw attention to the fact, already noted in
ref.~\cite{Cheung:2020djz}, that eq.~(\ref{NNSE}) may be rewritten as
\begin{equation}
  (\partial_0-\nu\nabla^2)u_i^a+\frac12 f^{abc}f_{ijk}u_j^b u_k^c=J_i^a,
  \label{NNSE2}
\end{equation}
where
\begin{equation}
  f_{ijk}v_j w_k=v_j\partial_j w_i-w_j\partial_j v_i
  \label{fijkfluid}
\end{equation}
for two arbitrary vectors $v_j$ and $w_j$. Recognising the components
of the Lie bracket of two vector fields (c.f. eq.~(\ref{V3def}) in the
relativistic case), we may instead write the field equation as
\begin{equation}
  (\partial_0-\nu\nabla^2){\bf u}+\frac12 [[{\bf u},{\bf u}]]={\bf J}.
  \label{NNSE3}
\end{equation}
We have here introduced the velocity field
\begin{equation}
  {\bf u}=u_i^a {\bf T}^a\partial_i
  \label{ubf}
\end{equation}
contracted with the appropriate generators. We have also introduced a
double bracket (c.f. eq.~(\ref{bracket})) consisting of simultaneous
Lie brackets in both the colour and diffeomorphism algebras:
\begin{equation}
  [[{\bf u},{\bf u}]]\equiv f^{abc}f_{ijk}u_j^b u_k^c.
  \label{bracket2}
\end{equation}
This theory fits into the scheme outlined in the rest of this paper
for Hamiltonian vector fields in Yang-Mills and abelian gauge theory
coupled to a scalar. That is, one may take the diffeomorphism algebra
associated with the linear theory of $u_j^a$, and use it to construct
a bracket for use in a cubic interaction term. As in the case of Chern-Simons theory, the Lie bracket of diffeomorphisms in the linearised theory survives in the
interaction term. Due to the solenoidal requirement, these are volume-preserving diffeomorphisms, which is directly analogous to the use of the Lorenz gauge in Chern-Simons theory. It is worthwhile to note that the presence of Lie brackets in fluid mechanics
has a direct physical interpretation: as we reviewed earlier, the Lie
bracket of two vector fields can be interpreted in terms of the Lie
derivative of one vector field along the other. In fluid mechanics,
Lie derivatives naturally arise from convection: the Lie
derivative of a vector $\vec{w}$ along a vector field $\vec{v}$
compares the change in the vector field $\vec{w}$ along the direction
of $\vec{v}$ with the form $\vec{w}$ would take if it were simply
dragged along the flow of $\vec{v}$. Thus, the kinematic algebra of
this theory is directly traceable to the convection properties of
fluid flows.

Another potential example of these ideas appears to be
ref.~\cite{Sheikh-Jabbari:2023eba}, which looks at a gauge theory for
shallow water waves first presented in ref.~\cite{Tong:2022gpg}. The
authors note that this description contains an area-preserving
diffeomorphism algebra. This is interpreted as a residual gauge
symmetry corresponding to the continuum limit of the freedom in
relabelling discrete elements of the two-dimensional surface of the
fluid. It would be interesting to study this theory in more detail, in
order to see if one can indeed intepret the area-preserving
diffeomorphisms as a kinematic algebra.

\section{Conclusion}
\label{sec:discuss}

Kinematic algebras are relatively new structures underlying gauge,
gravity and related field theories, and so remain somewhat
mysterious. For a general gauge theory, the kinematic algebra is
expected to be complicated, and not reducible to a straightforward Lie
algebra. However, in certain theories -- or sectors of theories -- we
are able to define a definite kinematic Lie algebra, without
necessarily understanding its precise origin.

In this paper, we have gained insights into this phenomenon by
studying simple abelian gauge theories. We argue that many known cases
of kinematic algebras for nonlinear (sub-)theories can be obtained by
taking well-defined subalgebras of the diffeomorphism algebra of gauge
(vector) fields. In the particular case of the symplectomorphism
subgroup, one may build a Poisson bracket involving the scalar field
entering the gauge field, and use this to generate interaction
terms. The case of self-dual Yang-Mills theory arises in this way, as
do its various generalisations that have been previously explored in
the literature. We obtain novel new examples of
kinematic algebras involving area-preserving diffeomorphism algebras,
such as the case of electromagnetism coupled to a complex scalar
field. Interestingly, we have not found examples of interacting theories containing Poisson brackets based on full real four-dimensional symplectomorphisms, and it would be interesting to know if such cases exist. 

We are also able to shed light on the issue of the gauge dependence of kinematic algebras, given that it is only for certain
gauge choices (even at abelian level) that the subgroup of
diffeomorphisms takes a minimal form.  Finally, we noted that Lorenz-gauge Chern-Simons theory~\cite{Ben-Shahar:2021zww} and the
non-abelian Navier-Stokes equation formulated in
ref.~\cite{Cheung:2020djz} are yet more cases in which the
diffeomorphism algebra already appearing at linear level can be used
to formulate a consistent interacting theory, thereby furnishing the
kinematic algebra with a direct geometric interpretation.

The study of kinematic algebras in recent years has provided tantalising hints of a hidden structure underlying gauge theories. Interestingly, making gauge invariance manifest makes for a simple Lagrangian, but obscures the kinematic algebra. On the other hand, making the kinematic algebra visible appears to lead to a much more complicated Lagrangian involving complex mathematical structures, thus obscuring the role of gauge invariance. We hope that our results may provide useful insights into how to navigate this quandary going forwards.

\section*{Acknowledgments}

We thank Leron Borsten, Graham Brown,  
Henrik Johansson, Ricardo Monteiro and Nicolas Pavao for helpful
discussions. This work has been supported by the UK Science and
Technology Facilities Council (STFC) Consolidated Grant ST/P000754/1
``String theory, gauge theory and duality''. KAW is supported by a
studentship from the UK Engineering and Physical Sciences Research
Council (EPSRC). S.N. is supported in part by STFC consolidated
grant T000708. SW is funded by the Knut and Alice Wallenberg Foundation grant KAW 2021.0170 and the Olle Engkvists Stiftelse grant 2180108.

\bibliography{refs}
\end{document}